\documentclass[aps,prl,reprint,groupedaddress]{revtex4-1}
\usepackage{amsmath,bm,mathrsfs,color}
\usepackage{graphicx}

\newcommand{\br}{\bm{r}}

\newcommand{\bpi}{\bm{\pi}}

\newcommand{\kp}{\bm{k}\cdot \bm{p}}
\newcommand{\D}{\Delta}

\newcommand{\wc}{\omega_{\rm c}}

\begin{document}

%\preprint{}

\title{%\underline{\small{\tt ver. 2.4 (\today)}}
%\vspace{5mm}\\
%Microscopic Theory of g-Factor in the Matrix Mechanics Representation and its Application to PbTe
Nonperturbative Matrix Mechanics Approach to Spin-Split Landau Levels and g-Factor in Spin-Orbit Coupled Solids
}


\author{Yuki Izaki}
\author{Yuki Fuseya}
\email[]{fuseya@uec.ac.jp}
%\thanks{}
\affiliation{Department of Engineering Science, University of Electro-Communications, Chofu, Tokyo 182-8585, Japan}


\date{\today}

\begin{abstract}
We have proposed a fully quantum approach to non-perturbatively calculate the spin-split Landau levels and g-factor of various spin-orbit coupled solids, based on the $\kp$ theory in the matrix mechanics representation. The new method considers the detailed band structure and the multiband effect of spin-orbit coupling irrespective of the magnetic field strength. An application of this method to PbTe, a typical Dirac electron system, is shown. Contrary to popular belief, it is shown that the spin-splitting parameter $M$, which is the ratio of the Zeeman to cyclotron energy, exhibits a remarkable magnetic-field-dependence. This field-dependence can rectify the existing discrepancy between experimental and theoretical results. We have also shown that $M$ evaluated from the fan diagram plot is different from that determined as the ratio of the Zeeman to cyclotron energy, which also overturns common belief.
\end{abstract}

%\keywords{}

\maketitle


%%%%%%%%%%%%%%%%%%%%%%%%%%%%%%%%%%%%%%%%%%%%%%%%

Change in the Zeeman splitting is the most direct observable consequence of spin-orbit coupling (SOC) in solids. The Zeeman energy, which is the energy difference between spin-up and down electrons under a magnetic field $B$, is usually defined as $E_Z = g\mu_B B$, where $g$ is the g-factor and $\mu_B$ is the Bohr magneton. For free electrons, $g=2$. On the contrary, for itinerant electrons in solids, it is modified due to the SOC and correlation between electrons \cite{Yafet1963,Shoenberg_book}. Particularly, large g-factors, such as in Zn ($g=170$) \cite{OSullivan1967} and Bi ($g=1060$) \cite{ZZhu2011}, are characteristics of the SOC mechanism with low carrier density. Therefore, measurements of the g-factor or the Zeeman energy can provide rare and valuable information of the SOC in solids.

Experimentally, the g-factor can be determined from quantum oscillations, where the frequency $F$ of an oscillation is given by $F/B=n+1/2 \pm M/2$ \cite{Shoenberg_book}. ($n$ is the Landau level index.) The spin-splitting parameter $M$ is defined as the ratio of the Zeeman energy to the cyclotron energy $E_c=\hbar \wc $. It is usually expressed as
\begin{align}
	M_{\rm zc}= \frac{E_Z}{E_c}=\frac{g\mu_B B}{\hbar \wc}=\frac{gm_c}{2m_e},
	\label{M-1}
\end{align}
where $m_e$ is the free electron mass, $m_c$ is the cyclotron mass and $\wc=eB/m_c$. $M_{\rm zc}$ characterizes the relative energy scale of the SOC to the kinetic energy in crystals \cite{Fuseya2015b}. 
When the SOC is negligibly small, it is expected to be $M_{{\rm zc}} \ll 1$.  $M_{\rm zc}$ increases as the impact of the SOC becomes significant. For $M_{{\rm zc}}=1$, the system is equivalent to the Dirac electrons \cite{Cohen1960,Wolff1964,Fuseya2015}. $M_{\rm zc}$ can also be greater than unity, depending on the contributions from the higher energy bands \cite{Kohler1975,Fauque2013,Orlita2015,Fuseya2015b}. 
In recent years, $M$ has attracted renewed interest since it is related to Berry's phase as $\phi_B=\pi M$, which is routinely discussed for topological materials \cite{DXiao2010,Ando2013,Murakawa2013,TLiang2017}.

Theoretically, on the other hand, it is extremely challenging to develop a fully quantum framework for the calculation of the g-factor and $M$, while considering the multiband effect of SOC. Yafet formulated perturbatively a basic idea of the g-factor \cite{Yafet1957,Yafet1963}. This was followed by several investigations on the g-factor, especially for semiconductors \cite{Roth1959,Cardona1963,Pidgeon1966,Mitchell1966}. Here, the specific symmetry of each crystal was analyzed while perturbatively considering the multiband (more than three bands) effect of SOC. 
For Luttinger Hamiltonian, which is the effective Hamiltonian for the two valence bands of Si or Ge, the spin-split Landau levels can be computed non-perturbatively using a specific symmetry operation \cite{Luttinger1956,Wallis1960,Yafet1963}. Thus far, the non-perturbative computation of the spin-split Landau levels has been limited to a two-band model, which is the minimum model for the interband effect \cite{Cohen1960,Wolff1964}. The multiband models have been solved using the perturbative approach only. 

Recently, a general analytic formula of g-factor was obtained on the basis of the relativistic multiband $\kp$ Hamiltonian with perturbation theories (L\"owdin partitioning) \cite{Fuseya2015b}. This method does not require analysis of the crystal, which makes it easily applicable to various solids. Although we were able to solve the half-a-century old mystery on the large anisotropic g-factor and $M_{\rm zc}$ of holes in Bi using this method, it is still a perturbative method. Further, the g-factor could only be obtained up to an order of $\mathcal{O} (\hbar \wc/\D) \sim \mathcal{O}(B^1)$, where $\D$ is half of the band gap. This is not sufficient for the cases with strong magnetic fields and narrow gaps, which have garnered a considerable interest recently, particularly for topological insulators, Weyl, and Dirac fermion systems. 
In fact, an issue related to the spin-splitting of PbTe, which is a typical narrow gap semiconductor, was reported.
Using the perturbative g-factor formula, it was predicted that $M_{\rm zc}=0.83$ \cite{Hayasaka2016}. On the contrary, $M=0.52-0.57$ experimentally, as obtained from the analysis of the Shubnikov-de Haas oscillation \cite{Akiba2018}. This large discrepancy cannot be attributed to theoretical error, and points towards a fundamental problem with the spin-splitting parameter that has not been encountered yet.

%========================================================================
\begin{figure}[tb]
	\includegraphics[width=8cm]{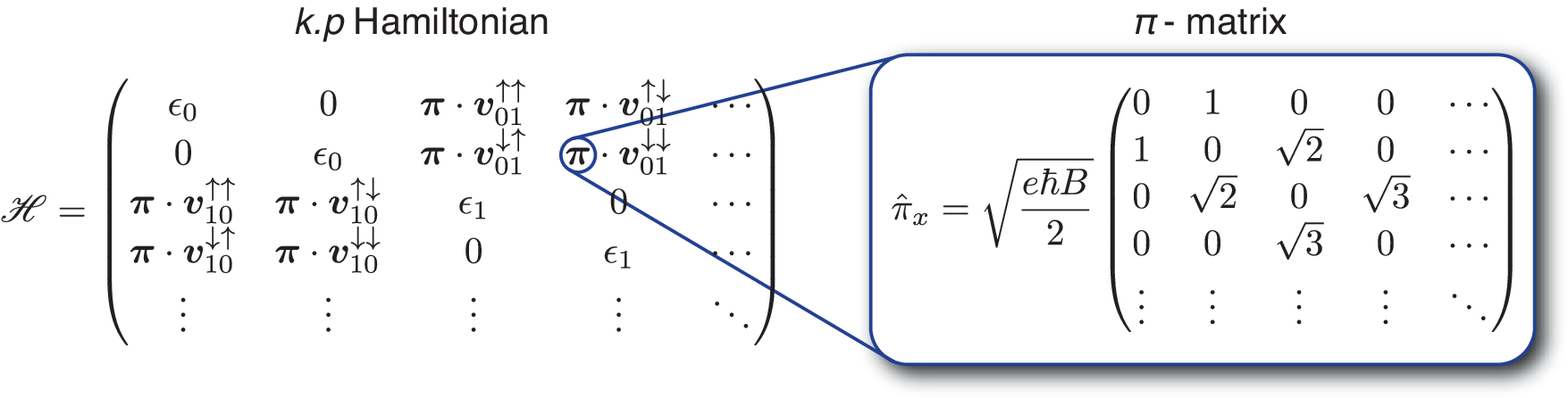}
	\caption{\label{Fig1} Schematic image of the $\pi$-matrix method. The kinematical momentum operator $\bpi$ in the $\kp$ Hamiltonian is expressed in the matrix mechanics representation.}
\end{figure}
%========================================================================

In this Letter, we propose a novel non-perturbative matrix mechanics approach to perform a rigorous calculation of the spin-split Landau levels and the g-factor, regardless of the magnitude of the field and the size of the gap. We have termed it the “$\pi$-matrix method” in this Letter (cf. Fig. \ref{Fig1}). It is based on a fully quantum theory without any semi-classical assumptions and the Bohr-Sommerfeld quantization rule. Here, we do not need any specific analysis of the crystal symmetry, whereas the previous methods \cite{Roth1959,Cardona1963,Pidgeon1966,Mitchell1966,Luttinger1956,Wallis1960} need an analysis of the specific symmetry of each crystal, so that the previous theory is unique for a particular crystal symmetry. This method can be easily combined with the band calculations, such as first principles calculations and tight-binding calculations. Consequently, the detailed band structures and the multiband effect of SOC can be taken into account automatically. 
To test the potential of the $\pi$-matrix method, we attempt to resolve a recently raised issue of discrepancy in $M$ of PbTe. We show that $M$ exhibits a remarkable dependence on magnetic fields, even though it has been believed to be invariant against the field [cf. Eq. \eqref{M-1}]. Our result can bridge the existing gap between theoretical and experimental studies. Further, we show that the spin-split parameter derived from the fan diagram plot is different from that defined as the ratio of the Zeeman to cyclotron energy.

%%%%%%%%%%%%%%%%%%%%%%%%%%%%%%%%%%%%%%%%%%%%%%%%

It was shown by Luttinger and Kohn \cite{Luttinger1955} that the motion of electrons in a periodic potential and a uniform magnetic field is described by the following equation, in the so-called Luttinger-Kohn representation \cite{Winkler_text,Voon_text,Fuseya2015}:
\begin{align}
	\sum_{\ell' \sigma'}\left[ \left( \epsilon_{\ell} + \frac{\pi^2}{2m_e}\right) \delta_{\ell \ell'}\delta_{\sigma \sigma'} +  \bpi\cdot \bm{v}_{\ell \ell'}^{\sigma \sigma'}\right]
	\psi_{\ell' \sigma'}(\br)=E\psi_{\ell \sigma}(\br).
	\label{kpeq}
\end{align}
$\bpi = -i\hbar \bm{\nabla} + e\bm{A}$ is the kinematical momentum operator under the magnetic field, where $\bm{A}$ is the vector potential and $e>0$ is the elementary charge. The wave function $\Psi(\br)$ can be expanded in terms of the band-edge Bloch functions $u_{\ell \sigma} (\br)$ as $\Psi (\br)= \sum_{\ell\sigma} \psi_{\ell \sigma} u_{\ell \sigma} (\br)$, where $\ell$ and $\sigma$ indicate the $\ell$-th Bloch band and its spin, respectively. It may be noted that $\sigma$ is not the bare spin, but expresses the degree of freedom of the Kramers doublet. $\epsilon_\ell$ is the band-edge energy of the $\ell$-th Bloch band. $\bm{v}_{\ell \ell'}^{\sigma \sigma'}$ is the matrix element of the velocity operator between $\psi_{\ell \sigma}^\dagger$ and $\psi_{\ell' \sigma'}$.
The multiband SOC effect is considered by $\bm{v}_{\ell \ell'}^{\sigma \sigma'}$ in a fully relativistic way, which is the strong merit of $\kp$ theory \cite{Luttinger1955}. In principle, the Hamiltonian in the Luttinger-Kohn representation can be written in a matrix form. For a $L$ band system, it is expressed in terms of a $2L \times 2L$ matrix. Initially, it seems that the energy of such electrons can be easily obtained by diagonalizing the Hamiltonian, which is true for the cases in the absence of magnetic field. However, this is not true in the presence of magnetic field since the commutation relation
\begin{align}
	\bpi \times \bpi = -i e\hbar \bm{B}
	\label{commutation}
\end{align}
cannot be satisfied by the simple diagonalization. This commutation relation prevents the rigorous calculation of the Landau levels. As discussed earlier, a possible way to circumvent this theoretical difficulty is the inclusion of the perturbation theory \cite{Yafet1957,Yafet1963,Roth1959,Cardona1963,Pidgeon1966,Mitchell1966,Fuseya2015b,Hayasaka2016}.

We use an unconventional yet simple idea to resolve this difficulty, which is the incorporation of matrix mechanics \cite{Heisenberg1925,Born1925}. Because the commutation relation Eq. \eqref{commutation} is essentially the same as that of the harmonic oscillation, $\bm{\pi}$ can be expressed in terms of a matrix  \cite{Born1925}. By replacing $\bpi$ in Eq. \eqref{kpeq} by the matrix form $\hat{\bpi}$, we obtain the Hamiltonian as the combination of the Luttinger-Kohn and the matrix mechanics representation (see Fig. \ref{Fig1}). The basis of the matrix $\hat{\bpi}$ is the Landau level index $n$ in the present case. With this matrix of $\hat{\bpi}$, the spin-split Landau levels can be calculated rigorously by a simple numerical diagonalization, even though the matrix size of the Hamiltonian becomes large, i.e. $(2LN)\times (2LN)$, where $N$ is the number of Landau levels considered. This method is gauge invariant, which guarantees the validity of the theory under a magnetic field, since the commutation relation Eq. \eqref{commutation} holds for arbitrary gauge. The parameters $\epsilon_\ell$ and $\bm{v}_{\ell \ell'}^{\sigma \sigma'}$ in the Hamiltonian can be directly calculated from the band calculations, such as tight-binding and first principles calculations, {\it at zero fields}. To the best of our knowledge, this is the first non-perturbative method that can be used to calculate the spin-split Landau levels while considering the detailed band structure and the multiband effect of SOC under any magnetic field strength. We verified the validity of the present method for the well-known case of free electrons and Dirac electrons. The details of the $\pi$-matrix method are given in the Supplemental Material \cite{SeeSM}.

%%%%%%%%%%%%%%%%%%%%%%%%%%%%%%%%%%%%%%%%%%%%%%%%
%========================================================================
\begin{figure}
	\includegraphics[width=7cm]{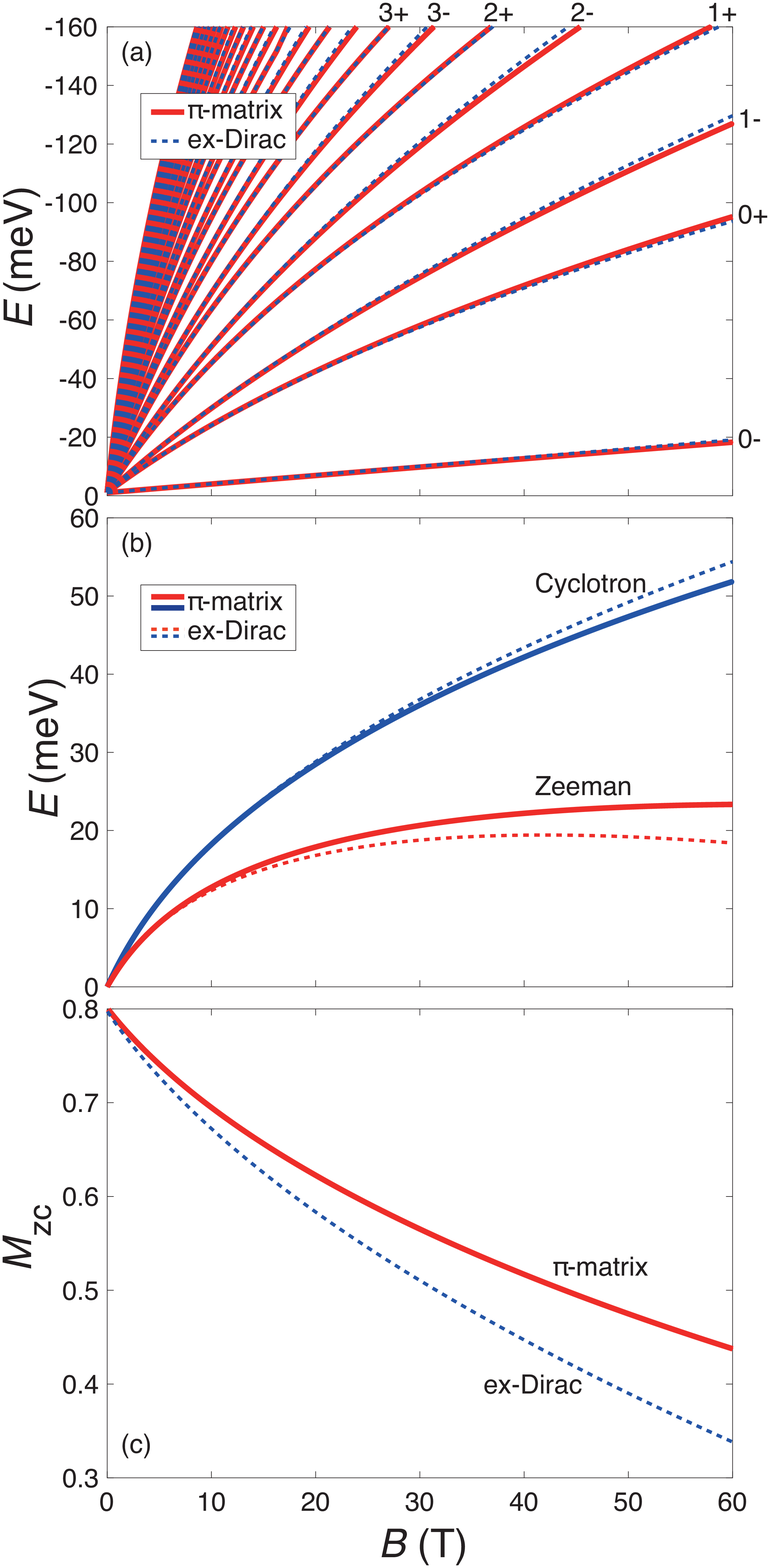}
	\caption{\label{Fig2} Magnetic-field-dependence of (a) spin-split Landau levels, (b) Zeeman $E_Z$ and cyclotron $E_c$ energy, and (c) spin-splitting parameter $M_{\rm zc}$. ($E_{Z, c}$ and $M_{\rm zc}$ are obtained for $n=2$.) The solid and dashed lines represent the results obtained by the $\pi$-matrix method and the extended Dirac model, respectively.}
\end{figure}
%======================================================================== 
We now apply the $\pi$-matrix method to PbTe, which is a narrow gap IV-VI semiconductor with a strong SOC. As mentioned earlier, an issue regarding large discrepancy between experimentally and theoretically reported values of $M$ in PbTe has been raised recently. 
To resolve this issue, we calculate the Zeeman energy, the cyclotron energy, and their ratio $M_{\rm zc}$, by the $\pi$-matrix method along with the relativistic tight-binding model by Lent {\it et al} \cite{Lent1986}. The results obtained agree with those obtained by another model by Lach-hab {\it et al}. \cite{Lachhab2000}. The details are given in \cite{SeeSM}.

Figure \ref{Fig2} (a) shows the spin-split Landau levels for the top valence band at the $L$-point as a function of the magnetic field. (The experimental values of $M$ were obtained for hole doped samples \cite{Akiba2018}.) The magnetic field is along the (001) direction, and the wavenumber along the field is set to be at the extremum of energy. The Landau level index $n$ and their spins $\sigma$ are uniquely determined from the weak-field-limit value of $M_{\rm zc}$ obtained by using the L\"owdin partitioning \cite{SeeSM}. Each energy level exhibits a sublinear field-dependence, suggesting that the system is close to the Dirac electrons, whose energy is given by $E_{j}^{\rm D}=\sqrt{\D^2 + 2\D \hbar \wc j}$, ($j=n+\sigma/2=0, 1, 2, \cdots$). For perfect Dirac electrons, the lowest Landau level is field invariant. However, the obtained lowest Landau level exhibits a weak field-dependence. This deviation from the perfect Dirac electrons is a clear indication of the contribution from the other bands, i.e., the multiband effect of SOC, which are ignored in the ordinary Dirac electron model. 
This multiband effect of SOC can be expressed by introducing the additional g-factor term up to the order of $\mathcal{O}(B^1)$ in the form
\begin{align}
	E_{n, \sigma}^{\rm exD}=\sqrt{\D^2 + 2\D \hbar \wc (n+\sigma/2)}+\frac{\sigma}{2}g' \mu_B B ,
	\label{exD}
\end{align}
which is called as the extended Dirac model \cite{Baraff1965,Vecchi1976,ZZhu2011,ZZhu2018}. (For holes, a negative sign is needed in Eq. \eqref{exD}.) To date, the value of $g'$ has been phenomenologically determined to fit the experimental data \cite{ZZhu2011,ZZhu2012,ZZhu2018}. We, for the first time, determine this value microscopically. The corresponding results for $g'=-10.4$ are shown in Fig. \ref{Fig2} (a) as dashed lines. (Note that the absolute value of the lowest Landau level $(n, \sigma)=(0, -)$ increases with the magnetic field when $g'<0$.) It is evident from Fig. \ref{Fig2} (a) that the spin-split Landau levels of PbTe can be well reproduced by the extended Dirac model, at least for $B<60$ T.

From the spin-split Landau levels, we evaluate the cyclotron energy $E_{c, n} = E_{n, +}-E_{n-1, +}$ and the Zeeman energy $E_{Z, n}=E_{n, +}-E_{n, -}$, which are shown in Fig. \ref{Fig2} (b) for $n=2$.
Both $E_{c, n}$ and $E_{Z, n}$ are proportional to $B^1$ in the weak field limit ($B\lesssim 2$ T). In the zero-field limit, $g=41.0$, which agrees with the previous value obtained perturbatively \cite{Hayasaka2016,SeeSM}. However, $E_{Z, c}$ exhibit a sublinear behavior in the strong field regime. Therefore, the g-factor, which is a coefficient of $B^1$, cannot be well-defined in this regime. This field-dependence is roughly explained by the extended Dirac model, as shown in Fig. \ref{Fig2} (b). The observed deviation of the extended Dirac model from the $\pi$-matrix method, which is particularly noticeable for $B \gtrsim 30$ T, indicates that the former is not adequate for this field region. These deviations arise because the higher order corrections, which are not considered in the extended Dirac model but rigorously considered in the $\pi$-matrix method, are not negligible at high fields. 
Next, we calculate the spin-splitting parameter $M_{{\rm zc}}=E_{Z}/E_{c}$ to investigate the long established assumption that $M_{{\rm zc}}$ is independent of the magnetic field. This is because, in Eq. \eqref{M-1}, both $E_c$ and $E_Z$ have been obtained up to only $\mathcal{O} (B^1)$ in the existing theories. Consequently, the field-dependence of $M_{{\rm zc}}$ has never been examined in detail.

However, we observe that $M_{\rm zc}$ exhibits a remarkable field-dependence, which is clear from Fig. \ref{Fig2} (c), where $n=2$. At zero field, $M_{\rm zc}=0.80$, and it rapidly decreases as $M_{\rm zc}=0.46$ at $B=$ 55 T. 
This drastic reduction is remarkable, and we found that it can be qualitatively explained using the extended Dirac model. From this model, it is easy to derive that 
\begin{align}
	M_{\rm zc}=\frac{E_{j+1}^{\rm D} - E_j^{\rm D}+g'\mu_B B}{E_{j+1}^{\rm D} - E_j^{\rm D}}.
\end{align}
The above equation clearly shows that $M_{\rm zc} < 1$ and it is a decreasing function of $B$ for $g'<0$. Therefore, the reduction of $M_{\rm zc}$ in Fig. \ref{Fig2} (c) is a logical consequence of the increase in the lowest Landau level in Fig. \ref{Fig2} (a).

%========================================================================
\begin{figure}
	\includegraphics[width=7cm]{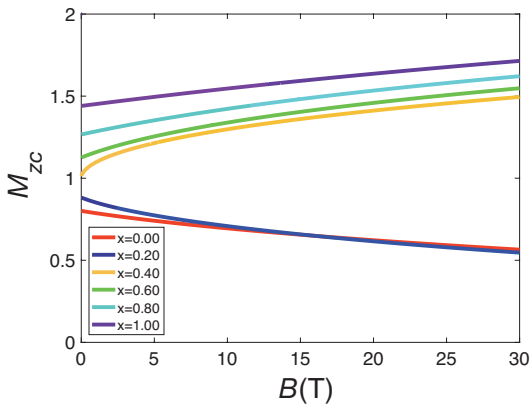}
	\caption{\label{Fig3} Spin-splitting parameter $M_{\rm zc}$ for Pb$_{1-x}$Sn$_x$Te. The band inversion (therefore the topological transition) occurs at $x=0.38$.}
\end{figure}
%========================================================================
It is expected that $M_{\rm zc}$ should be an increasing function of $B$ for $g'>0$. This can be verified by substituting Sn for Pb, i.e., Pb$_{1-x}$Sn$_x$Te. It is well known that the band inversion between the conduction and valence bands occurs at around $x\simeq 0.4$ \cite{Dimmock1971}, which accompanies the topological transition from trivial to non-trivial \cite{Hsieh2012}. In the Lent {\it et al.} model, the inversion point is at $x=0.38$, where $M_{\rm zc}$ changes from $M_{\rm zc}<1$ to $M_{\rm zc}>1$ \cite{Hayasaka2016,SeeSM}. 
Figure \ref{Fig3} shows the field-dependence of $M_{\rm zc}$ for Pb$_{1-x}$Sn$_x$Te using the Lent {\it et al.} model. As expected, $M_{\rm zc}$ is an increasing function of $B$ after the inversion at $x = 0.38$, since $M_{\rm zc}>1$ so that $g'>0$ for $x>0.38$. These results are entirely consistent with the understandings obtained from the extended Dirac model.

%\section{Fan diagram}
%========================================================================
\begin{figure}
	\includegraphics[width=7cm]{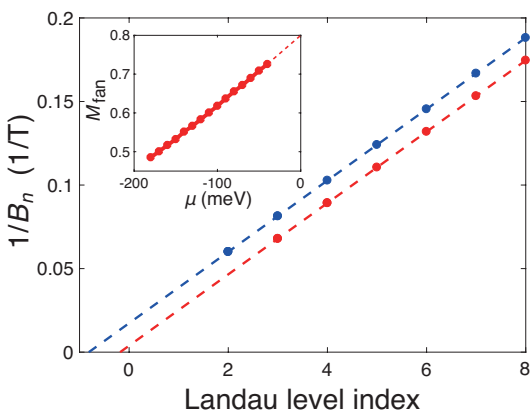}
	\caption{\label{Fig4} Plot of the reciprocal fields $1/B_n$ for PbTe, where $\mu$ touches the bottom of the $n$-th Landau level. Here $M_{\rm fan}=x_- - x_+ = 0.63$. The inset shows the $\mu$-dependence of $M_{\rm fan}$.}
\end{figure}
%========================================================================
An alternate method for evaluating $M$ is the fan diagram plot \cite{Shoenberg_book}, where $1/B_n$ is plotted as a function of the Landau level index. Here, $B_n$ is the field at which the chemical potential touches the $n$-th Landau level. These reciprocal fields are linearly fitted for each spin, following which $M$ is evaluated from the difference of $x$-intercept, i.e., $M_{\rm fan}=x_- - x_+$. In principle, $M_{\rm fan}$ is not exactly equal to $M_{\rm zc}$. The significant difference can be easily understood from the fact that $M_{\rm fan}$ is independent of $B$, but dependent on $\mu$, and vice versa is true for $M_{\rm zc}$. The relation $M_{\rm fan}=M_{\rm zc}$ is only true for the cases of free electrons with Zeeman energy and Dirac electrons \cite{SeeSM}. 
Figure \ref{Fig4} shows the fan diagram of PbTe for the Lent {\it et al.} model with $\mu=-92$ meV from the top of the valence band (the hole density is $n_h=3.7 \times 10^{18}$ cm$^{-3}$). We obtain $M_{\rm fan}^{\rm Lent}=0.63$ by fitting the calculated $1/B_n$. The $\mu$-dependence of $M_{\rm fan}$ is shown in the inset of Fig. \ref{Fig4}. $M_{\rm fan}$ tends to approach the zero-field value of $M_{\rm zc} =0.80$ in the $\mu=0$ limit. This is also consistent with the analysis using the extended Dirac model \cite{SeeSM}.

Experimentally, $M_{\rm zc}$ can be determined directly from the Zeeman and cyclotron energy \cite{Fuseya2015b}. It is not appropriate to compare the approximate value of $M$ obtained from the Lifshitz-Kosevich formula with the present theoretical value, since this formula is not based on a fully quantum theory. According to Akiba {\it et al.}  \cite{Akiba2018}, $M_{\rm fan}^{\rm exp}=0.56$ (evaluated from Fig. 9), which is less than that obtained using the Lent {\it et al.} model $M_{\rm fan}^{\rm Lent}=0.63$. Although the agreement is not perfect yet, we have successfully removed the large discrepancy between experiment and theory. Further enhancements of the theoretical accuracy can be achieved by improving the accuracy of the band calculations, since the value of $M$ is quite sensitive to the details of the band structure. Even the high-energy bands (more than 1 eV far from the conduction band) can change $M$ from $M_{\rm fan}^{\rm Lent}=0.63$ to $M_{\rm fan}^{\rm Lach-hab}=0.40$ \cite{SeeSM}.

%%%%%%%%%%%%%%%%%%%%%%%%%%%%%%%%%%%%%%%%%%%%%%%%

In summary, we have demonstrated a novel non-perturbative method, which is based on matrix mechanics, for calculating the spin-split Landau levels. This is the first method to elucidate the following properties of the spin-splitting parameter $M$. (i) $M_{\rm zc}$ is largely dependent on the magnetic field; (ii) In general, $M_{\rm zc}$ is not equivalent to $M_{\rm fan}$. The origin of these previously unknown properties is the multiband effect of SOC and the higher order corrections in $B$. This method also provides an explanation for the large discrepancy between experimentally and theoretically reported values of $M$ in PbTe. 
Apart from the specific case of PbTe considered here, this method can be useful for other systems in which the SOC plays a relevant role, such as topological insulators ($M\sim 2$ in Bi$_2$Se$_3$ has not been explained yet \cite{Kohler1975,Fauque2013,Orlita2015}), Weyl, and Dirac fermion systems.

\begin{acknowledgments}
	We thank M. Tokunaga and K. Akiba for helpful discussions. This work is supported by JSPS KAKENHI (grants No. 16K05437 and 19H01850).
\end{acknowledgments}

\bibliography{Bismuth.bib,PbTe.bib,footnote.bib}

%\begin{acknowledgments}
%\end{acknowledgments}

%\bibliography{Bismuth.bib}

\end{document}